\documentclass[a4paper,12pt]{article}

\pdfoutput=1
\usepackage{jheppub}
\usepackage[T1]{fontenc}

\usepackage{graphicx}
\usepackage{cancel}
\usepackage{amssymb}
\usepackage{textcomp}
\usepackage{amsmath}
\usepackage{bm}
\usepackage{times}
\usepackage{epsfig}
\usepackage{epstopdf}
\usepackage{color}
\usepackage{multirow}

\def\lsim{\mathrel{\raise.3ex\hbox{$<$\kern-.75em\lower1ex\hbox{$\sim$}}}}
\def\gsim{\mathrel{\raise.3ex\hbox{$>$\kern-.75em\lower1ex\hbox{$\sim$}}}}

\def\beq{\begin{equation}}
\def\eeq{\end{equation}}
\def\be{\begin{equation}}
\def\ee{\end{equation}}
\def\bea{\begin{eqnarray}}
\def\eea{\end{eqnarray}}

\def\to{\rightarrow}

\title{Simplified dark matter models in the light of AMS-02 antiproton data}

\author{Tong Li}

\emailAdd{tong.li@monash.edu}
\affiliation{
ARC Centre of Excellence for Particle Physics at the Tera-scale, School of Physics and Astronomy, Monash University, Melbourne, Victoria 3800 Australia}

\abstract{
In this work we perform an analysis of the recent AMS-02 antiproton flux and the antiproton-to-proton ratio in the framework of simplified dark matter models.  
To predict the AMS-02 observables we adopt the propagation and injection parameters determined by the observed fluxes of nuclei. We assume that the dark matter particle is a Dirac fermionic dark matter, with leptophobic pseudoscalar or axialvector mediator that couples only to Standard Model quarks and dark matter particles. We find that the AMS-02 observations are consistent with the dark matter hypothesis within the uncertainties. The antiproton data prefer a dark matter (mediator) mass in the 700 GeV--5 TeV region for the annihilation with pseudoscalar mediator and greater than 700 GeV (200 GeV--1 TeV) for the annihilation with axialvector mediator, respectively, at about 68\% confidence level. The AMS-02 data require an effective dark matter annihilation cross section in the region of $1 \times 10^{-25}$ -- $1 \times 10^{-24}$ ($1 \times 10^{-25}$ -- $4 \times 10^{-24}$) ${\rm cm}^3/{\rm s}$ for the simplified model with pseudoscalar (axialvector) mediator. The constraints from the LHC and Fermi-LAT are also discussed.
}

\begin{document}

\maketitle
\flushbottom
\newpage

\section{Introduction}

Charged cosmic rays connect information about galactic astrophysics with that about possibly new fundamental particle physics.
Explaining the precise measurements of cosmic ray spectra requires the detailed knowing the propagation and injection of cosmic rays and the microscopic properties of the fundamental particle such as dark matter. The recent observations of cosmic ray nuclei by AMS-02, e.g. proton~\cite{Aguilar:2015ooa}, antiproton~\cite{Aguilar:2016kjl}, Helium~\cite{Aguilar:2015ctt}, etc., provide updated understanding the propagation/source injection parameters and leptophobic dark matter models. These measurements gain attentions of both astrophysicists and particle physicists~\cite{Stref:2016uzb,Cuoco:2016eej,Cui:2016ppb,Feng:2016loc,Huang:2016tfo,Liu:2016gyv,Abe:2016wck,Lin:2016ezz}.

The propagation parameters can be determined by fitting the secondary-to-primary ratio of cosmic ray nuclei, such as the Boron-to-Carbon ratio (B/C), and the ratio of secondary nuclei, such as the Beryllium isotope ratio $\rm ^{10}Be/^{9}Be$. The observed proton flux can further fix the unified injection parameters of all nuclei. Based on these obtained parameters, one can derive an up-to-date astrophysical background for the secondary production of antiprotons so as to study the extra sources like dark matter. A self-consistent
way to take into account the dark matter source is to propagate the antiproton spectrum induced by dark matter annihilation through the Galaxy and calculate the antiproton flux under the exact same set of the above astrophysical parameters. This procedure ensures a consistent astrophysical treatment of cosmic rays from the standard astrophysical source and dark matter~\cite{Balazs:2015iwa}.

In this work, we examine the constraint of AMS-02 data of antiproton flux and antiproton-to-proton ratio on leptophobic simplified models of dark matter. This hypothesis is widely adopted in the analysis of dark matter search at the Large Hadron Collider (LHC), satellites in the sky and underground direct detection experiments~\cite{Buchmueller:2013dya,Arina:2014yna,Alves:2015pea,Abdallah:2015ter,Boveia:2016mrp,Arina:2016cqj}. It uses minimal and general theoretical assumptions with only two parameters, i.e. the dark matter mass and the mediator mass, and the simultaneous presence of various annihilation channels provides the dark matter models with considerable flexibility. We specifically consider a Dirac fermionic dark matter, with pseudoscalar and axialvector mediators that couple only to quarks and dark matter particles. The resulted dark matter annihilations are not velocity suppressed~\cite{Kumar:2013iva}. Meanwhile the dark matter-nucleon elastic scattering cross sections are spin-dependent (SD) thus do not receive stringent constraint from direct detection. We also derive the AMS-02 preferred region in the parameter space of the dark matter models.


This paper is organized as follows. In Sec.~\ref{sec:Propagation} we describe the propagation equation and injection spectra for cosmic ray nuclei.  The values of corresponding parameters are also given.
In Sec.~\ref{sec:Models}, we describe the simplified dark matter models we use.
Our numerical results are given in Sec.~\ref{sec:Results}.
Finally, in Sec.~\ref{sec:Concl} we summarize our conclusions.

\section{Propagation and Injection of Cosmic Rays}
\label{sec:Propagation}

Cosmic rays in the Galaxy are categorized into primary and secondary types \cite{1964ocr..book.....G, Blandford:1987pw, Stawarz:2009ig, Aharonian:2011da}.  The interstellar mediums (ISM) are accelerated to produce primary cosmic rays. The produced primary cosmic ray protons and nuclei collide with the ISM and then produce secondary cosmic rays.
The cosmic ray propagation within the galaxy is described by the following transport equation~\cite{Strong:2007nh}
\begin{eqnarray}
{\partial \psi\over \partial t}&=&Q(\vec{r},p)+\vec{\nabla}\cdot \left(D_{xx}\vec{\nabla}\psi-\vec{V}\psi\right)+{\partial\over \partial p}p^2D_{pp}{\partial\over \partial p}{1\over p^2}\psi \nonumber\\
&&-{\partial\over \partial p}\left[\dot{p}\psi-{p\over 3}\left(\vec{\nabla}\cdot \vec{V}\right)\psi\right]-{\psi\over \tau_f}-{\psi\over \tau_r},
\label{propagation}
\end{eqnarray}
where $\psi(\vec{r},t,p)$ is the density of cosmic rays per unit of total particle momentum $p$. $\vec{V}$ is the convection velocity and $\tau_f (\tau_r)$ is the time scale for fragmentation (radioactive decay). The spatial diffusion coefficient is usually written in this form
\begin{eqnarray}
D_{xx}=\beta D_0 (R/R_0)^\delta ,
\end{eqnarray}
with $R$ and $\beta$ being the rigidity and particle velocity divided by light speed respectively.  The diffusion coefficient in momentum space, i.e. $D_{pp}$, is dependent on the square of the Alfven velocity $v_A$.  $z_0$ is the height of the cylindrical diffusion halo.  The above key propagation parameters can be constrained by fitting the latest ratios of nuclei, that is the Boron-to-Carbon ratio ($\rm B/C$) and the Beryllium ratio ($\rm ^{10}Be/^{9}Be$).  We adopt the diffusion reacceleration model and the values of propagation parameters shown in Table \ref{tab:parameter}, determined by the $\rm B/C$ and $\rm ^{10}Be/^{9}Be$ data~\cite{Cui:2016ppb}.

In Eq.~(\ref{propagation}), the source term of cosmic ray species $i$ can be described by the product of the spatial distribution and the injection spectrum function
\begin{eqnarray}
Q_i(\vec{r},p)=f(r,z)q_i(p) .
\end{eqnarray}
For the spatial distribution of the primary cosmic rays we use the following supernova remnants distribution
\begin{eqnarray}
f(r,z)=f_0\left({r\over r_\odot}\right)^a{\rm exp}\left(-b \ {r-r_\odot\over r_\odot}\right){\rm exp}\left(-{|z|\over z_s}\right),
\label{snr}
\end{eqnarray}
where $r_\odot=8.5 \ {\rm kpc}$ is the distance between the Sun and the Galactic center, the height of the Galactic disk is $z_s=0.2 \ {\rm kpc}$, and the two parameters $a$ and $b$ are chosen to be 1.25 and 3.56, respectively~\cite{Lin:2014vja}. 
We assume the following power law with one break for the injection spectrum of various nuclei
\begin{eqnarray}
q_i&\propto& \left\{
                \begin{array}{ll}
                  \left(R/R_{\rm br}^p\right)^{-\nu_1}, & R\leq R_{\rm br}^p \\
                  \left(R/R_{\rm br}^p\right)^{-\nu_2}, & R> R_{\rm br}^p
                \end{array}
              \right. \ .
\label{injection}
\end{eqnarray}
The corresponding injection parameters in Eq.~(\ref{injection}), i.e. rigidity break $R_{\rm br}^p$ and power law indexes $\nu_1,\nu_2$, can be determined by fitting the latest AMS-02 proton data~\cite{Aguilar:2015ooa}.  We adopt injection parameters obtained by performing such a fit in Ref.~\cite{Cui:2016ppb}.  The values of these injection parameters are shown in Table~\ref{tab:parameter}, together with the Fisk potential $\phi_i \ (i = p, \bar{p})$ for solar modulation effect.




\begin{table}[h]
\begin{center}
\resizebox{15cm}{!} {
\begin{tabular}{|c|c|c|c|c|c|c|c|}
        \hline
        propagation & value && nucleon injection & value && solar modulation & value\\
        \hline
        $D_0 \ (10^{28} \ {\rm cm}^2 \ {\rm s}^{-1})$ & 7.09 && $\nu_1$ & 1.702 && $\phi_{p} \ ({\rm MV})$ & 550\\
        \hline
        $\delta$ & 0.349 && $\nu_2$ & 2.399 && $\phi_{\bar{p}} \ ({\rm MV})$ & 400\\
        \hline
        $R_0 \ ({\rm GV})$ & 4 && $R_{\rm br}^p$ \ ({\rm GV}) & 11.48 && $-$ & $-$\\
        \hline
        $v_A \ ({\rm km} \ {\rm s}^{-1})$ & 38.14 && $A_p$ (see caption) & 4.325 && $-$ & $-$\\
        \hline
        $z_0$ \ ({\rm kpc}) & 5.47 &&    $-$    & $-$ && $-$ & $-$ \\
        \hline
\end{tabular}}
\end{center}
\caption{Parameters of propagation, nucleon injection and solar modulation and their values adopted in our numerical analysis.  The proton flux is normalized to  $A_p$ at 100 GeV in the units of $10^{-9} \ {\rm cm}^{-2} \ {\rm s}^{-1} \ {\rm sr}^{-1} \ {\rm MeV}^{-1}$.}
\label{tab:parameter}
\end{table}

\section{The Simplified Dark Matter Models}
\label{sec:Models}

In this section, we describe the simplified dark matter models restricted by the AMS-02 data of antiproton flux and antiproton-to-proton ratio.
We assume that dark matter is composed of Dirac fermionic particles, which we denote by $\chi$. The dark matter particles couple to the Standard Model (SM) quarks through a pseudoscalar mediator $S$ or an axialvector mediator $V$. The corresponding interactions are as follows~\cite{Abdallah:2015ter}
\begin{eqnarray}
{\cal L}_{\rm pseudoscalar} &=& -ig_{\rm DM}^S S\bar{\chi}\gamma_5 \chi - ig_q^S S\sum_{q=u,d,s,c,b,t}{m_q\over v_0}\bar{q}\gamma_5 q,
\label{eq:interaction}\\
{\cal L}_{\rm axialvector} &=& -g_{\rm DM}^A V_\mu\bar{\chi}\gamma^\mu\gamma_5 \chi - g_q^A V_\mu\sum_{q=u,d,s,c,b,t}\bar{q}\gamma^\mu\gamma_5 q,
\label{eq:interaction0}
\end{eqnarray}
where $v_0=246$ GeV. Following the general choices in the analysis of dark matter searches in literatures, we take $g_{\rm DM}^S=g_q^S=1$ and $g_{\rm DM}^A=1, g_q^A={1\over 4}$ in the calculations below. Under the above assumptions the dark matter models are described by two parameters, i.e. the dark matter mass $m_\chi$ and the mediator mass $m_S$ or $m_V$. The scan range for these parameters is
\begin{eqnarray}
5 \ {\rm GeV} < m_\chi, m_S, m_V < 10 \ {\rm TeV}.
\end{eqnarray}

Induced by the interactions in Eqs.~(\ref{eq:interaction}) and (\ref{eq:interaction0}), the pairs of dark matter particle $\chi$ can either annihilate into SM quark pairs via the mediator particle in s channel $\bar{\chi}\chi \to S/V\to \bar{q}q$, or annihilate into the mediator pairs in t channel followed by mediators decaying to SM quarks $\bar{\chi}\chi \to SS/VV \to \bar{q}q\bar{q}'q'$. The resulting cosmic ray spectra can thus be categorized into 2-body spectrum and 4-body spectrum, respectively.

The source term arising from dark matter annihilation contributing to the cosmic ray species $i$ is given by
\begin{eqnarray}
Q_i^\chi(r,p)=\frac{\rho_\chi^2(r)\langle \sigma_{\rm ann} v\rangle}{2 m_\chi^2}\frac{dN_i}{dE},
\label{dmsource}
\end{eqnarray}
where $\langle \sigma_{\rm ann} v\rangle$ is the total velocity averaged dark matter annihilation cross section of all kinematically allowed channels. $dN_i/ dE$ is the total energy spectrum of cosmic ray particle $i$ produced in the annihilation, that is the sum of 2-body spectrum and 4-body spectrum $dN_i/ dE=(dN_i/ dE)_{\rm 2-body}+(dN_i/ dE)_{\rm 4-body}$.

For the 2-body spectrum, one has
\begin{eqnarray}
\left({dN_i\over dE}\right)_{\rm 2-body} &=& \sum_q {\langle \sigma_{\rm ann} v\rangle_q\over \langle \sigma_{\rm ann} v\rangle} \frac{dN_i^q}{dE}+{\langle \sigma_{\rm ann} v\rangle_g\over \langle \sigma_{\rm ann} v\rangle} \frac{dN_i^g}{dE},
\label{2body}
\end{eqnarray}
where $\langle \sigma_{\rm ann} v\rangle_q =\sigma_{\rm ann} v(\bar{\chi}\chi\to S/V\to q\bar{q})$, $\langle \sigma_{\rm ann} v\rangle_g =\sigma_{\rm ann} v(\bar{\chi}\chi\to S\to gg)$ for the pseudoscalar mediator case and $\langle \sigma_{\rm ann} v\rangle_g =0$ for the axialvector mediator case. $dN_i^q/dE$ and $dN_i^g/dE$ are the cosmic ray spectra given by dark matter direct annihilating into quark pairs $\bar{q}q$ and gluons $gg$, respectively.
The 4-body spectrum is
\begin{eqnarray}
\left({dN_i\over dE}\right)_{\rm 4-body} &=& \sum_q {\langle \sigma_{\rm ann} v\rangle_{\rm Med}\over \langle \sigma_{\rm ann} v\rangle} {\Gamma_{{\rm Med}\to q\bar{q}}\over \Gamma_{\rm Med}}\frac{d\bar{N}_i^q}{dE}+{\langle \sigma_{\rm ann} v\rangle_{\rm Med}\over \langle \sigma_{\rm ann} v\rangle} {\Gamma_{{\rm Med}\to gg}\over \Gamma_{\rm Med}}\frac{d\bar{N}_i^g}{dE},
\label{4body}
\end{eqnarray}
where $\langle \sigma_{\rm ann} v\rangle_{\rm Med}=\sigma_{\rm ann} v(\bar{\chi}\chi\to SS/VV)$, $\Gamma_{{\rm Med}\to q\bar{q}}=\Gamma_{S/V\to q\bar{q}}$ and the total decay width of the mediator is $\Gamma_{\rm Med}=\Gamma_{S/V}$. $\Gamma_{{\rm Med}\to gg}=\Gamma_{S\to gg}$ for the pseudoscalar mediator case and $\Gamma_{{\rm Med}\to gg}=0$ for the axialvector mediator case. $d\bar{N}_i^q/dE$ and $d\bar{N}_i^g/dE$ are the cosmic ray spectra in the lab frame given by the spectrum from the mediator decay in its rest frame, denoted by $dN_i^q/dE_0$ and $dN_i^g/dE_0$, after a Lorentz boost~\cite{Elor:2015tva,Elor:2015bho}:
\begin{eqnarray}
{d\bar{N}_i^{q,g}\over dE}&=&2\int^{t_{1,\rm max}}_{t_{1,\rm min}}{dx_0\over x_0\sqrt{1-\epsilon^2}}{dN_i^{q,g}\over dE_0},
\end{eqnarray}
where
\begin{eqnarray}
t_{1,\rm max}&=&{\rm min}\left[1,{2x\over \epsilon^2}\left(1+\sqrt{1-\epsilon^2}\right)\right],\\
t_{1,\rm min}&=&{2x\over \epsilon^2}\left(1-\sqrt{1-\epsilon^2}\right)
\end{eqnarray}
with $\epsilon=m_{\rm Med}/m_\chi$ and $x=E/m_\chi\leq 0.5$. The expressions of dark matter annihilation cross sections and mediator decay widths in Eqs.~(\ref{2body}) and (\ref{4body}) are collected in Appendix. As a result of the non-trivial involvement of the mediator, $\langle \sigma_{\rm ann} v\rangle$ and $dN_i/ dE$ are dependent on both the dark matter mass and the mediator mass. AMS-02 data thus play an important role in constraining these two parameters.

\begin{figure}[t]
\begin{center}
\includegraphics[scale=1,width=7cm]{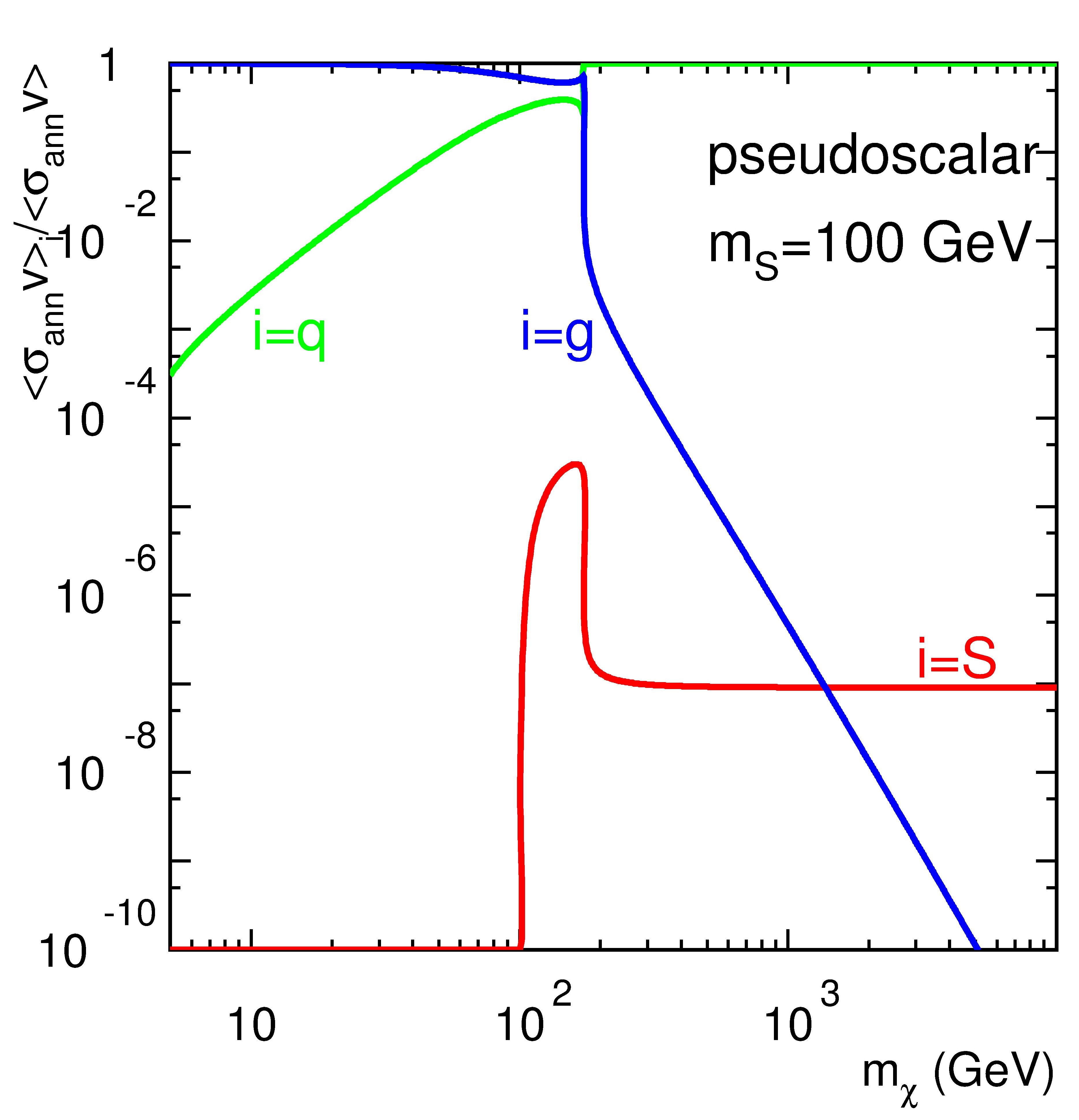}
\includegraphics[scale=1,width=7cm]{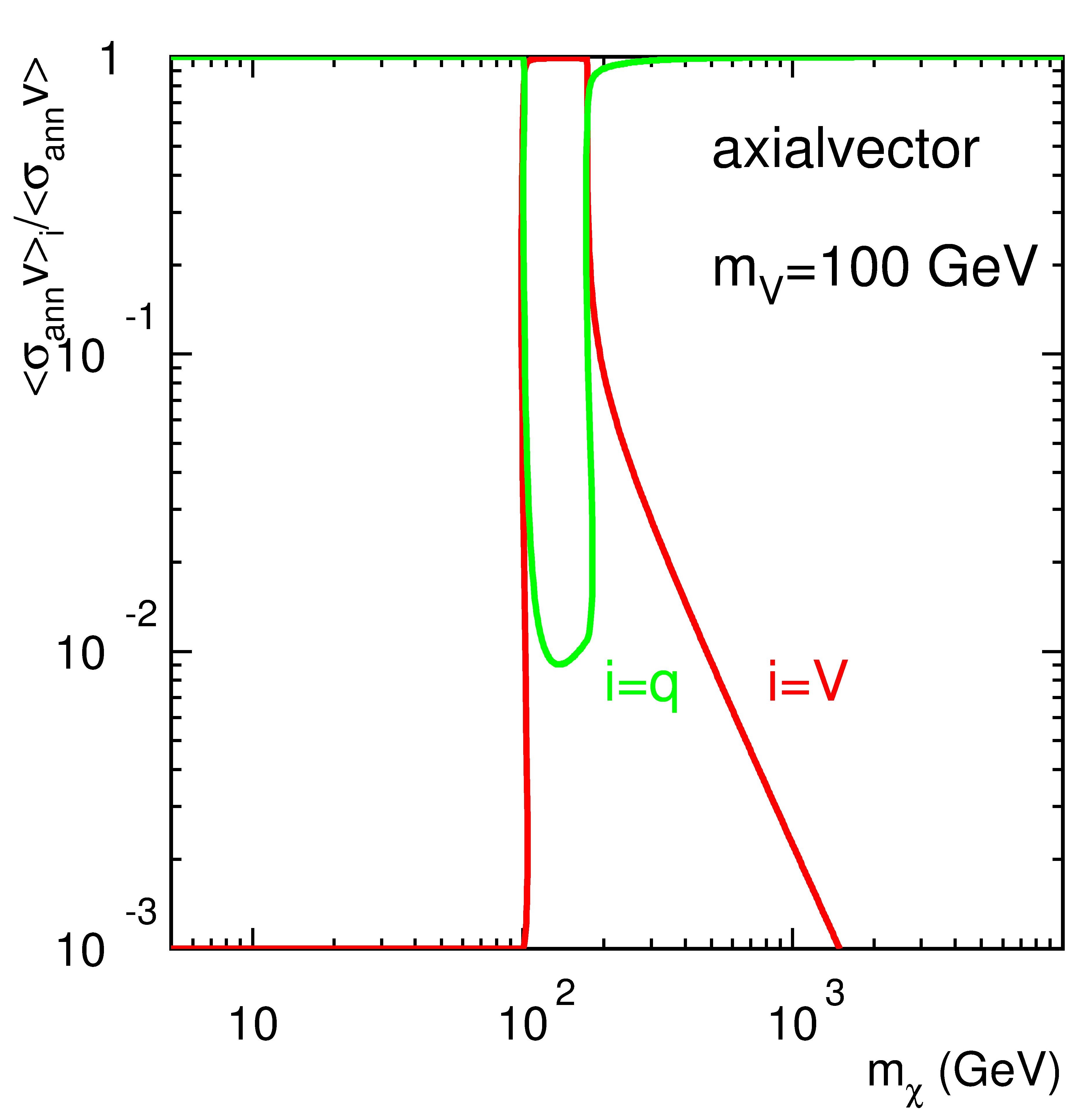}
\end{center}
\caption{The annihilation cross section fractions $\langle \sigma_{\rm ann} v\rangle_i/ \langle \sigma_{\rm ann} v\rangle$ as a function of $m_\chi$ for the pseudoscalar mediator case (left) and the axialvector mediator case (right). The mediator mass is fixed to be 100 GeV.
}
\label{anni}
\end{figure}

\begin{figure}[t]
\begin{center}
\includegraphics[scale=1,width=7cm]{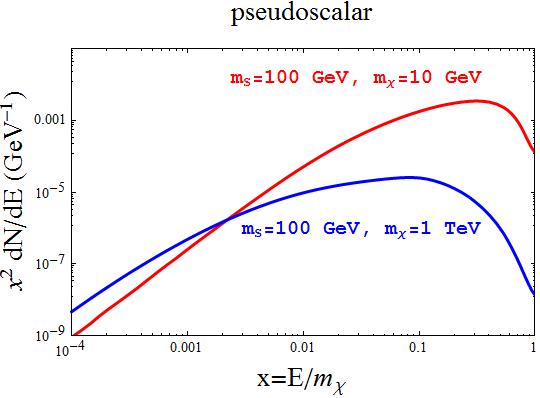}
\includegraphics[scale=1,width=7cm]{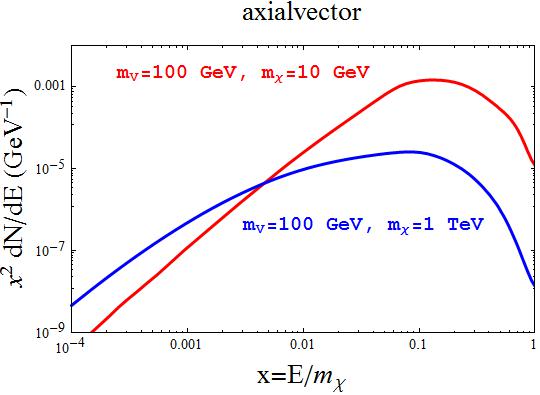}
\end{center}
\caption{Total antiproton spectrum $x^2dN_i/dE$ as a function of $x=E/m_\chi$ for the pseudoscalar mediator case (left) and the axialvector mediator case (right).
}
\label{dnde}
\end{figure}

We show the $\langle \sigma_{\rm ann} v\rangle_i/ \langle \sigma_{\rm ann} v\rangle$ as a function of $m_\chi$ in Fig.~\ref{anni}. For pseudoscalar mediator case, we find that $\bar{\chi}\chi\to gg$ channel is dominant for small dark matter mass region. After $t\bar{t}$ channel is open, $\bar{\chi}\chi\to \bar{q}q$ channel turns to be dominant. $\bar{\chi}\chi\to SS$ channel is always very small as it is a process through p wave. For axialvector mediator case, $\bar{\chi}\chi\to \bar{q}q$ is dominant before $\bar{\chi}\chi\to VV$ is forbidden and after $\bar{\chi}\chi\to \bar{t}t$ is open. In Fig.~\ref{dnde} we show the resulted total antiproton spectrum $x^2dN_i/dE$ as a function of $x=E/m_\chi$.


We use a generalized Navarro-Frenk-White (NFW) profile to describe dark matter spatial distribution~\cite{NFW}
\begin{eqnarray}
\rho_\chi(r)=\rho_0\frac{(r/r_s)^{-\gamma}}{(1+r/r_s)^{3-\gamma}}.
\end{eqnarray}
Here
the coefficient $\rho_0$ is $0.26 \ {\rm GeV/cm^3}$ and
the radius of the galactic diffusion disk is $r_s=20$ kpc.
We fix the inner slope of the halo profile as $\gamma=1$.

\section{Results}
\label{sec:Results}
As discussed in Sec.~\ref{sec:Propagation}, the propagation and injection parameters of cosmic rays are determined by fitting the $\rm B/C$ and $\rm ^{10}Be/^{9}Be$ data and cosmic ray proton data from AMS-02, respectively.  The parameters in Table~\ref{tab:parameter} thus imply prediction for cosmic ray measurements inferred from standard astrophysical sources.  One can investigate the constraint on extra sources, such as dark matter, based on this fiducial astrophysical background.

For each group of dark matter mass and mediator mass, we use PPPC4DMID~\cite{Cirelli:2010xx} to generate the antiproton spectrum in Eqs.~(\ref{2body}) and (\ref{4body}), and calculate the dark matter annihilation cross sections following the formulas in Appendix. These dark matter model dependent variables are then passed into the public code Galprop v54~\cite{Strong:1998pw, Moskalenko:2001ya, Strong:2001fu, Moskalenko:2002yx, Ptuskin:2005ax} to ensure that near Earth cosmic ray fluxes from dark matter annihilation and background spectra are obtained in a consistent way.

The calculated cosmic ray fluxes, together with the measured data points, are put into a composite likelihood function, defined as
\begin{eqnarray}
- 2\ln{\cal L}= \sum_i {(f_i^{\rm th}-f_i^{\rm exp})^2\over \sigma_i^2} .
\label{sum}
\end{eqnarray}
Here $f_i^{\rm th}$ are the theoretical predictions and $f_i^{\rm exp}$ are the corresponding central value of the experimental data.   The uncertainty $\sigma_i$ combines the theoretical and experimental uncertainties in quadrature.  We stipulate a 50\% uncertainty of the theoretical prediction of antiproton flux and antiproton-to-proton ratio according to the estimates in Refs.~\cite{Cui:2016ppb,Trotta:2010mx, Auchettl:2011wi, Giesen:2015ufa}.  This uncertainty takes into account, amongst other, the uncertainty related to the fixed propagation parameters.
The sum in Eq.~(\ref{sum}) runs over all the AMS-02 antiproton cosmic ray spectral data points: the antiproton flux (57 points) and antiproton-proton ratio (57 points).

As the dark matter-nucleon elastic scattering cross sections induced by the simplified models we consider are spin-dependent, the most stringent constraints come from collider search and indirect detection of dark matter~\cite{CMS-PAS-EXO-16-037,Khachatryan:2016ecr,Sirunyan:2016iap,Karwin:2016tsw}. LHC performed dark matter search using events with large missing transverse momentum plus energetic jets~\cite{CMS-PAS-EXO-16-037} and dijet events~\cite{Khachatryan:2016ecr,Sirunyan:2016iap} at 13 TeV collisions. Their exclusion limits can be directly presented in the plane of dark matter mass vs. mediator mass for simplified model with a pseudoscalar mediator or an axialvector mediator. Moreover, Fermi Large Area Telescope (LAT) searched for gamma ray emission from Milky Way satellite galaxies using 6 years of data. They recently released the observed constraints on the dark matter annihilation cross section for pure $b\bar{b}$ channel~\cite{Fermi-LAT:2016uux}. We can convert the Fermi-LAT limit into a bound on our dark matter annihilation cross section. Suppose the $b\bar{b}$ component of the total annihilation cross section fixed by dark matter mass and mediator mass satisfies
\begin{eqnarray}
\langle \sigma_{\rm ann} v\rangle > \langle \sigma v\rangle_{bb}^{\rm Fermi-limit} {\langle \sigma_{\rm ann} v\rangle\over \langle \sigma_{\rm ann} v\rangle_b},
\label{Fermi}
\end{eqnarray}
we claim the corresponding set of $m_\chi, m_{\rm Med}$ is excluded.



\begin{figure}[t]
\begin{center}
\includegraphics[scale=1,width=7cm]{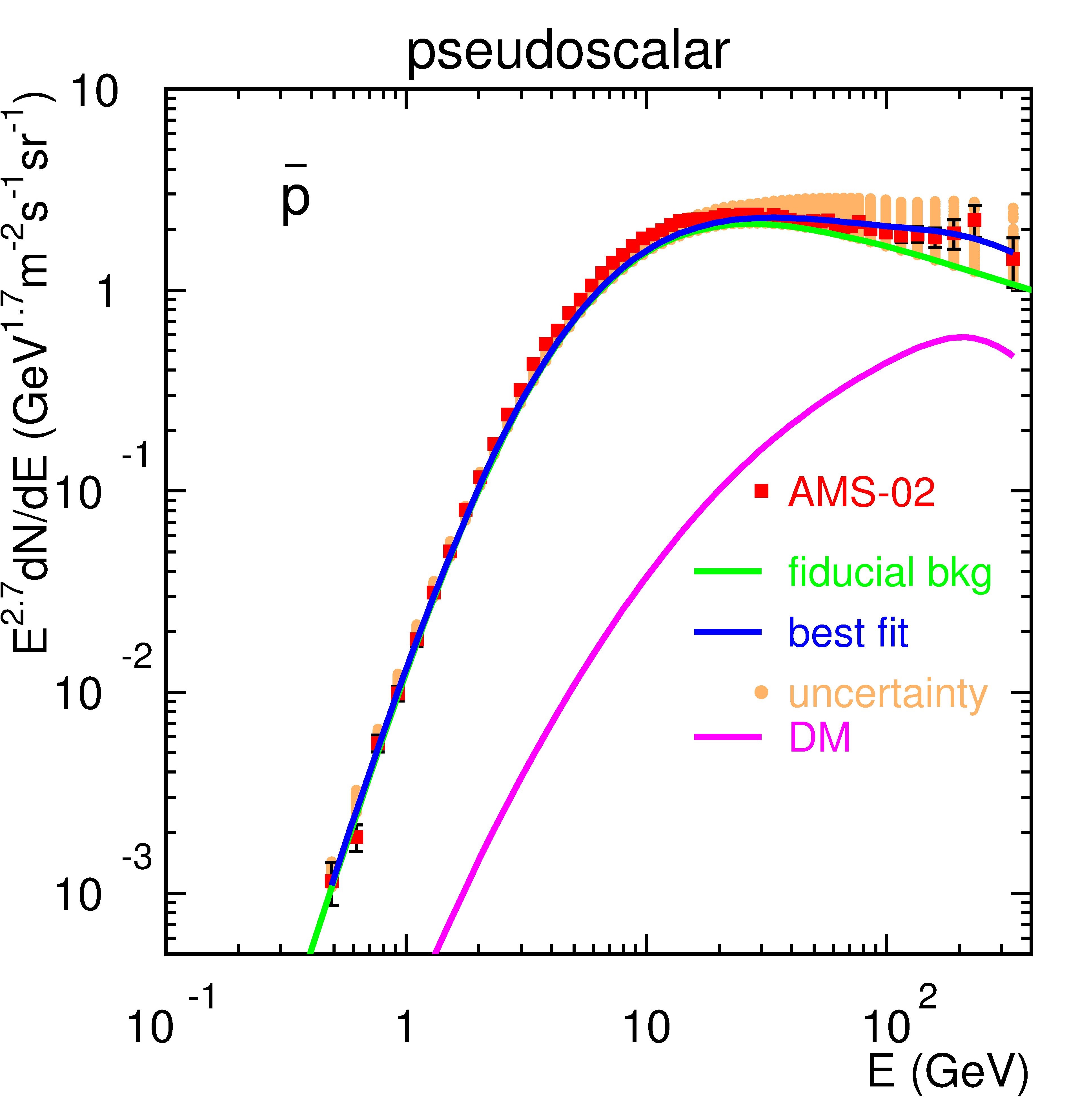}
\includegraphics[scale=1,width=7cm]{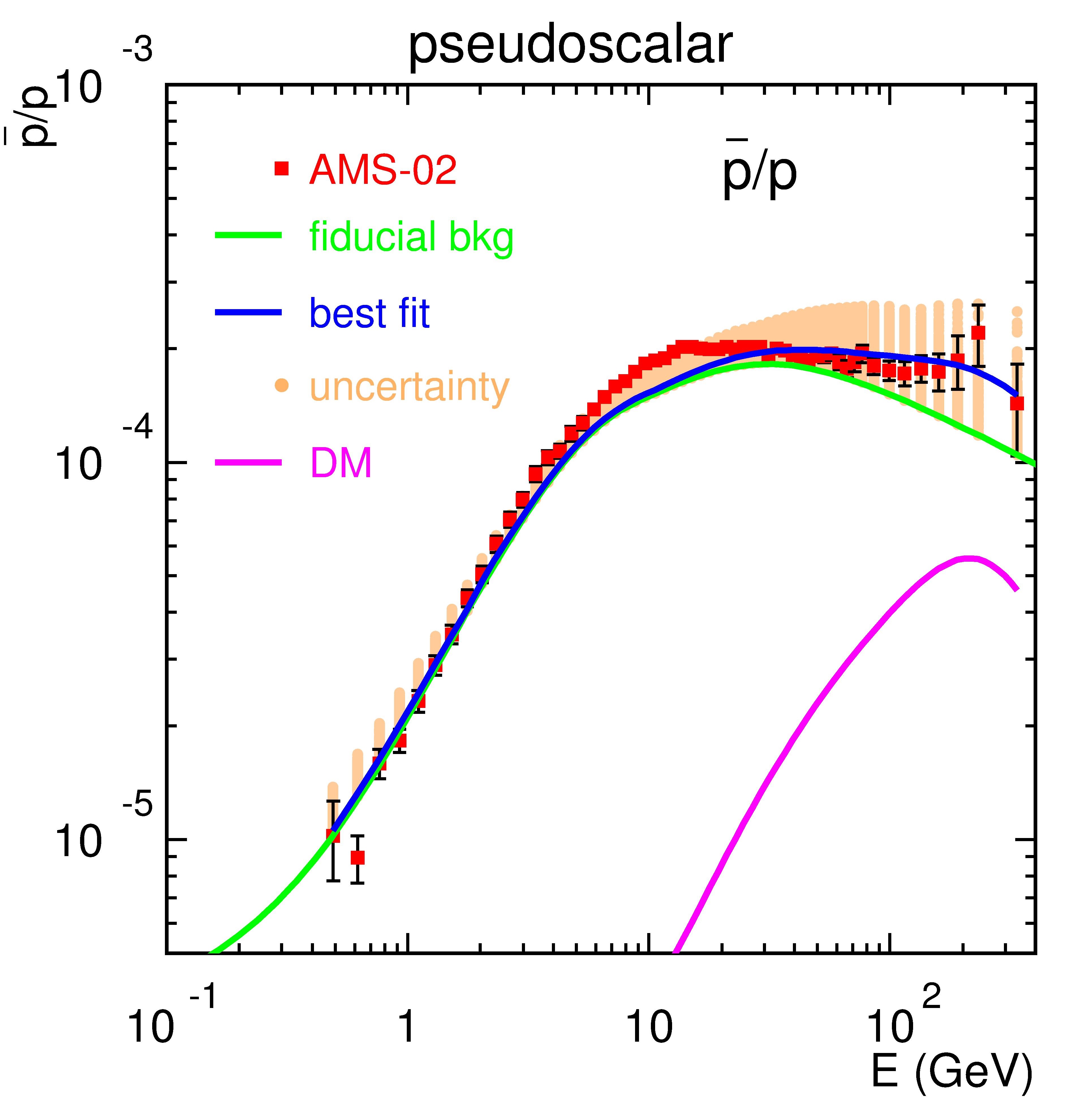}
\end{center}
\caption{Antiproton flux (left) and antiproton-to-proton ratio (right) observed by AMS-02 (red dots and dark error bars) in the simplified dark matter model with a pseudoscalar mediator.  The blue solid line shows the prediction of the total cosmic ray flux with dark matter parameter values that best fit the AMS-02 data. The total predicted flux is the sum of the background flux (green curve) and the dark matter contribution (purple curve).  Salmon dots indicate the $2\sigma$ confidence region of the prediction. 
}
\label{fig:fit1}
\end{figure}

\begin{figure}[t]
\begin{center}
\includegraphics[scale=1,width=7cm]{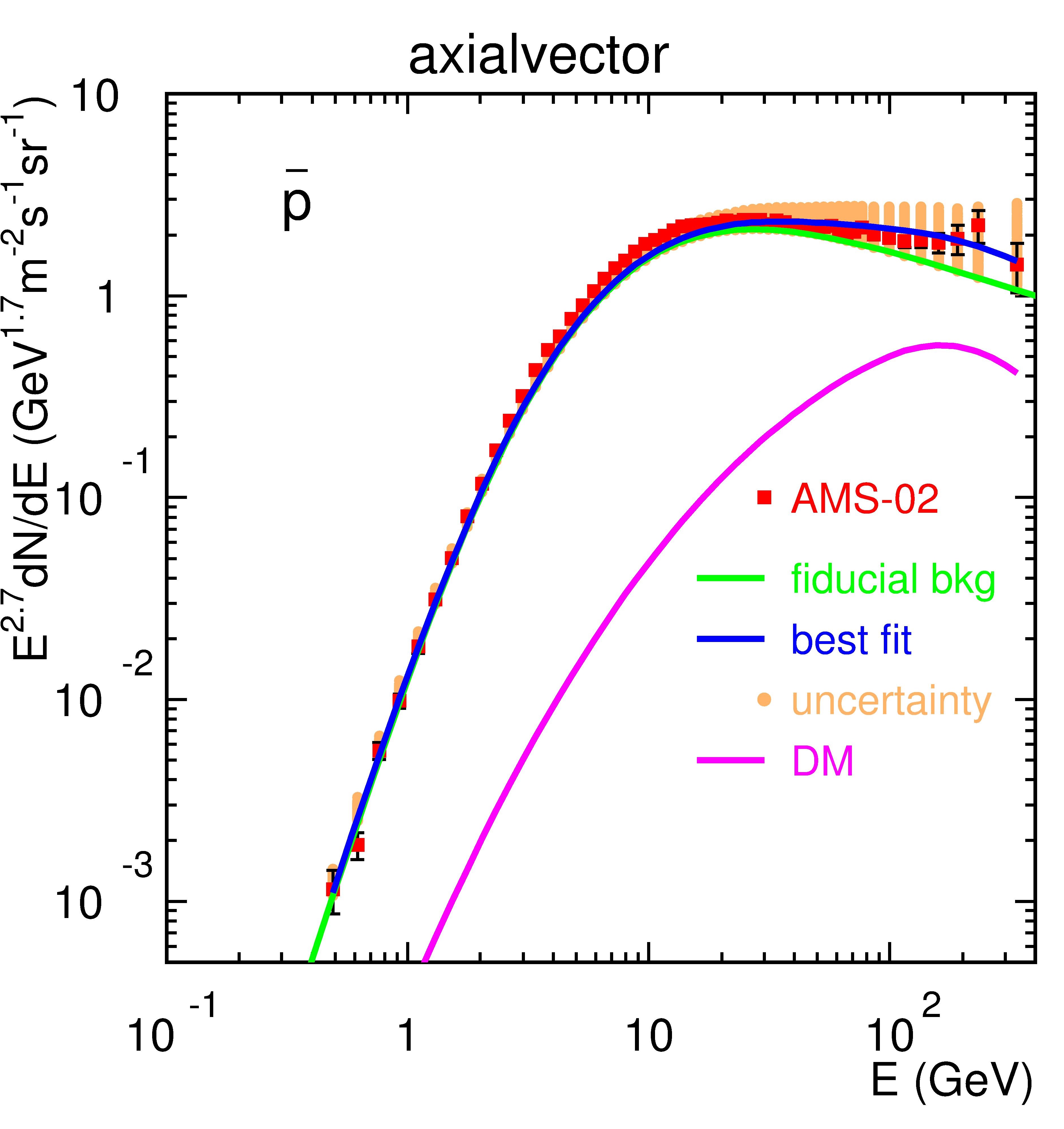}
\includegraphics[scale=1,width=7cm]{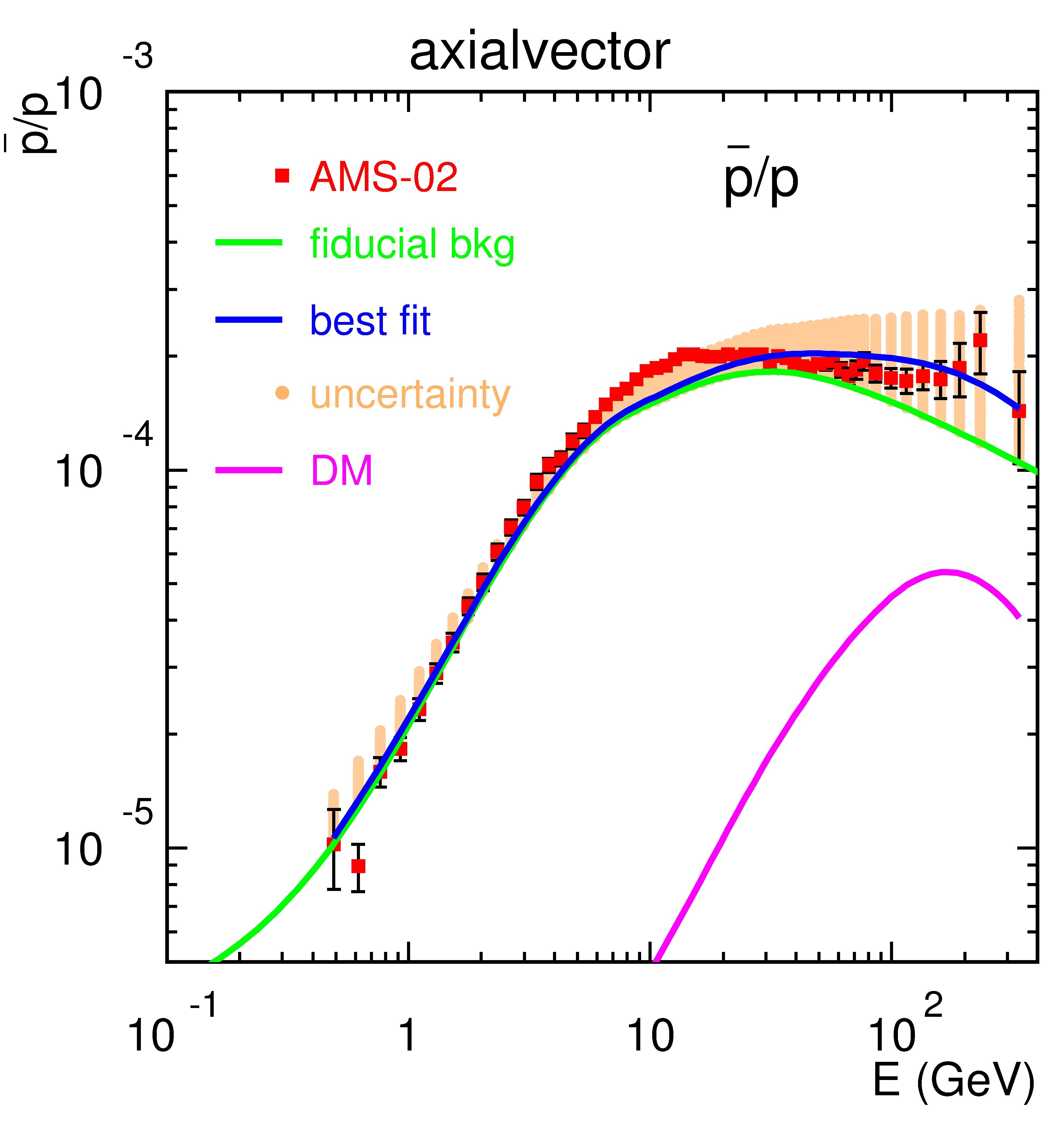}
\end{center}
\caption{Antiproton flux (left) and antiproton-to-proton ratio (right) observed by AMS-02 (red dots and dark error bars) in the simplified dark matter model with an axialvector mediator.
}
\label{fig:fit2}
\end{figure}

FIGs.~\ref{fig:fit1} and \ref{fig:fit2} show our main results: AMS-02 cosmic ray flux data are consistent with the dark matter hypothesis within the uncertainties.  The two plots in each figure display the antiproton cosmic ray: antiproton flux and antiproton-to-proton ratio.  AMS-02 central value measurements are shown by red dots and error bars in black indicate measurement uncertainties.  The green solid curves are obtained using the parameters shown in Table~\ref{tab:parameter} and display the predicted background flux originating from standard astrophysical sources.  The blue solid lines show the predictions of the total cosmic ray flux with dark matter contribution that fit the AMS-02 data best and are the sum of the background flux (green curve) and the dark matter contribution at the best fit point (purple curve).  A series of salmon colored vertical bars indicate the theoretical uncertainty of the dark matter prediction given by the $2\sigma$ confidence region of dark matter model parameters. The plots show that adding a dark matter contribution to the background flux yields a better fit to the AMS-02 data.
%
%

\begin{figure}[t]
\begin{center}
\includegraphics[scale=1,width=7.5cm]{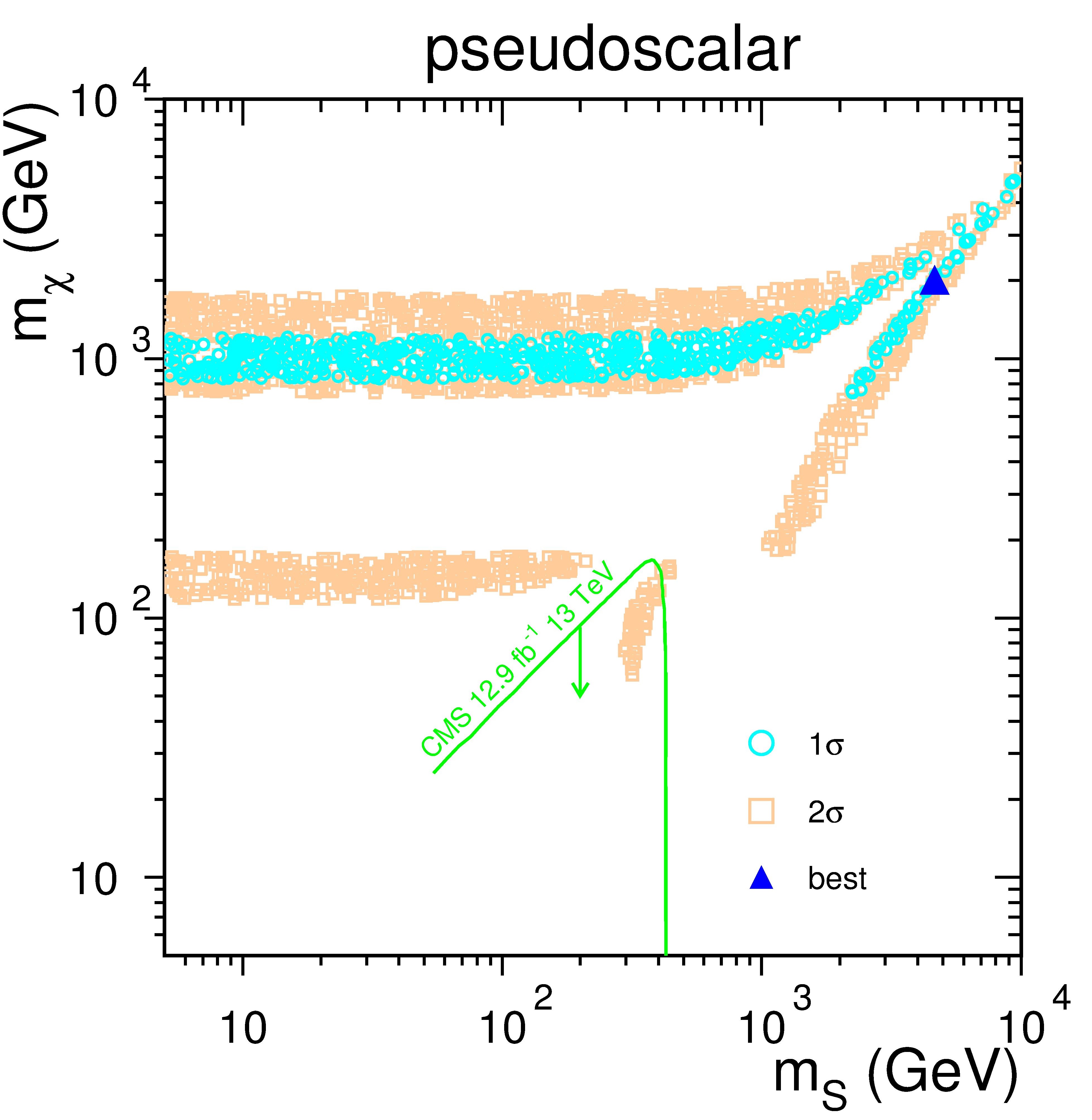}
\includegraphics[scale=1,width=7.5cm]{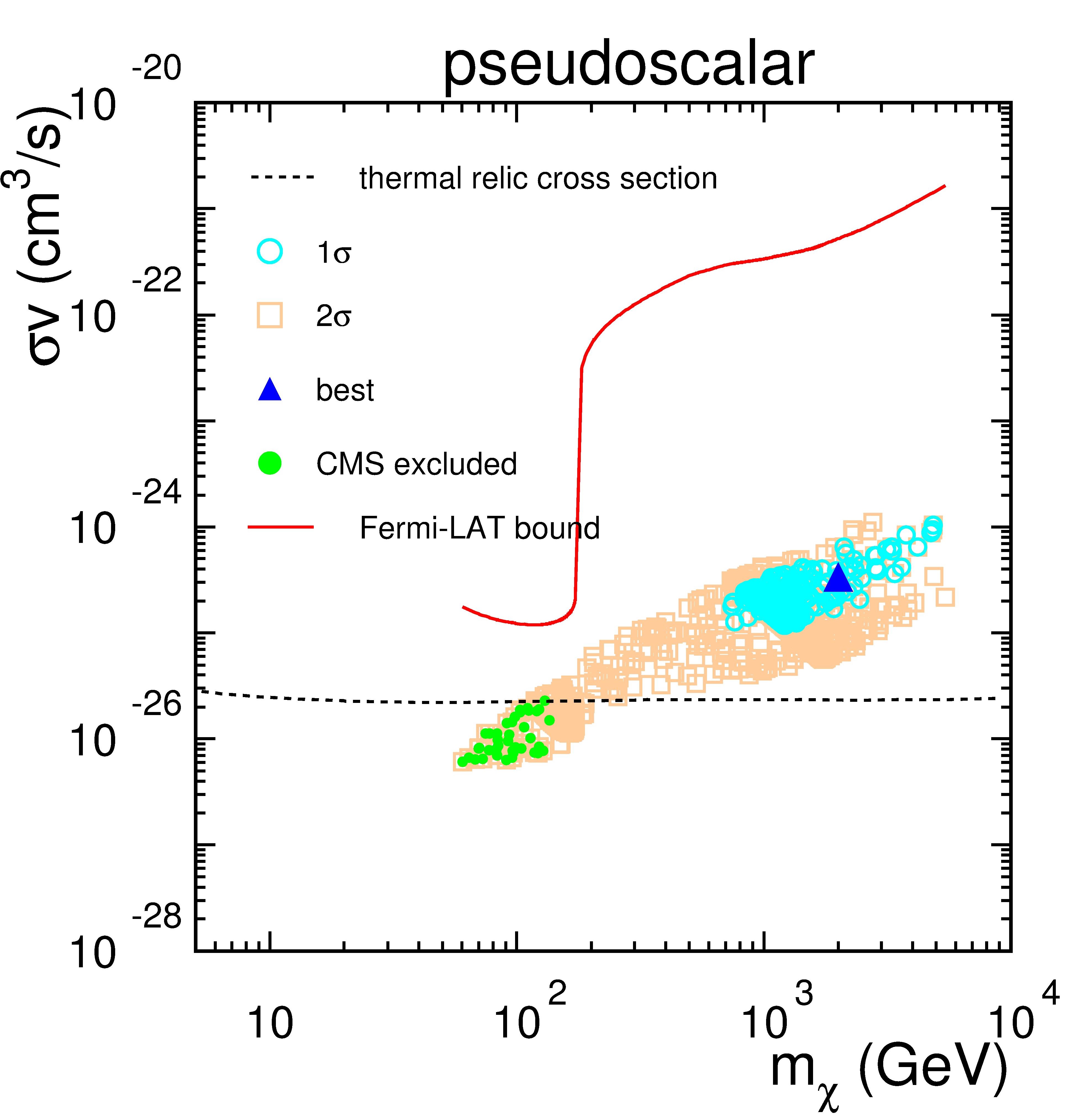}
\end{center}
\caption{Left: the AMS-02 favored region of masses ($m_\chi$ vs. $m_S$) in the simplified dark matter model with a pseudoscalar mediator we consider. The solid circles and squares estimate $1\sigma$ and $2\sigma$ confidence regions, respectively. The best fit point is indicated by a triangle. The green curve is the LHC exclusion limit~\cite{CMS-PAS-EXO-16-037}.  Right: the AMS-02 favored region of cross sections ($\sigma v$ vs. $m_\chi$). The green points are excluded by LHC search. The red curve is the converted upper bound from Fermi-LAT, i.e. the right hand side of Eq.~(\ref{Fermi}). The black dashed curve corresponds to the thermal cross section~\cite{Steigman:2012nb}.}
\label{fig:region1}
\end{figure}

In the left frame of Fig.~\ref{fig:region1} we show the regions of the mass parameter space preferred by the AMS-02 data and the LHC limit for the pseudoscalar mediator case.  Solid circles and squares denote the estimated $1\sigma$ and $2\sigma$ confidence regions, respectively. We find the AMS-02 antiproton data favor region $700~{\rm GeV}\lesssim m_\chi\lesssim 5~{\rm TeV}$ at about $1\sigma$ confidence level. The LHC excludes a part of the $2\sigma$ confidence region with $m_\chi\lesssim 170$ GeV and $300 \ {\rm GeV}\lesssim m_S\lesssim 420 \ {\rm GeV}$.


The right frame of Fig.~\ref{fig:region1} shows that the AMS-02 data require an effective dark matter annihilation cross section in the region of $1 \times 10^{-25}$ -- $1 \times 10^{-24}$ ($5 \times 10^{-27}$ -- $2 \times 10^{-24}$) ${\rm cm}^3/{\rm s}$ at about 1(2)$\sigma$ C.L. The LHC excludes a part of the region below thermal relic cross section, denoted by green dots. The Fermi-LAT bound becomes rather weak after $t\bar{t}$ channel is open and thus does not constrain the AMS-02 favored region.

\begin{figure}[t]
\begin{center}
\includegraphics[scale=1,width=7.5cm]{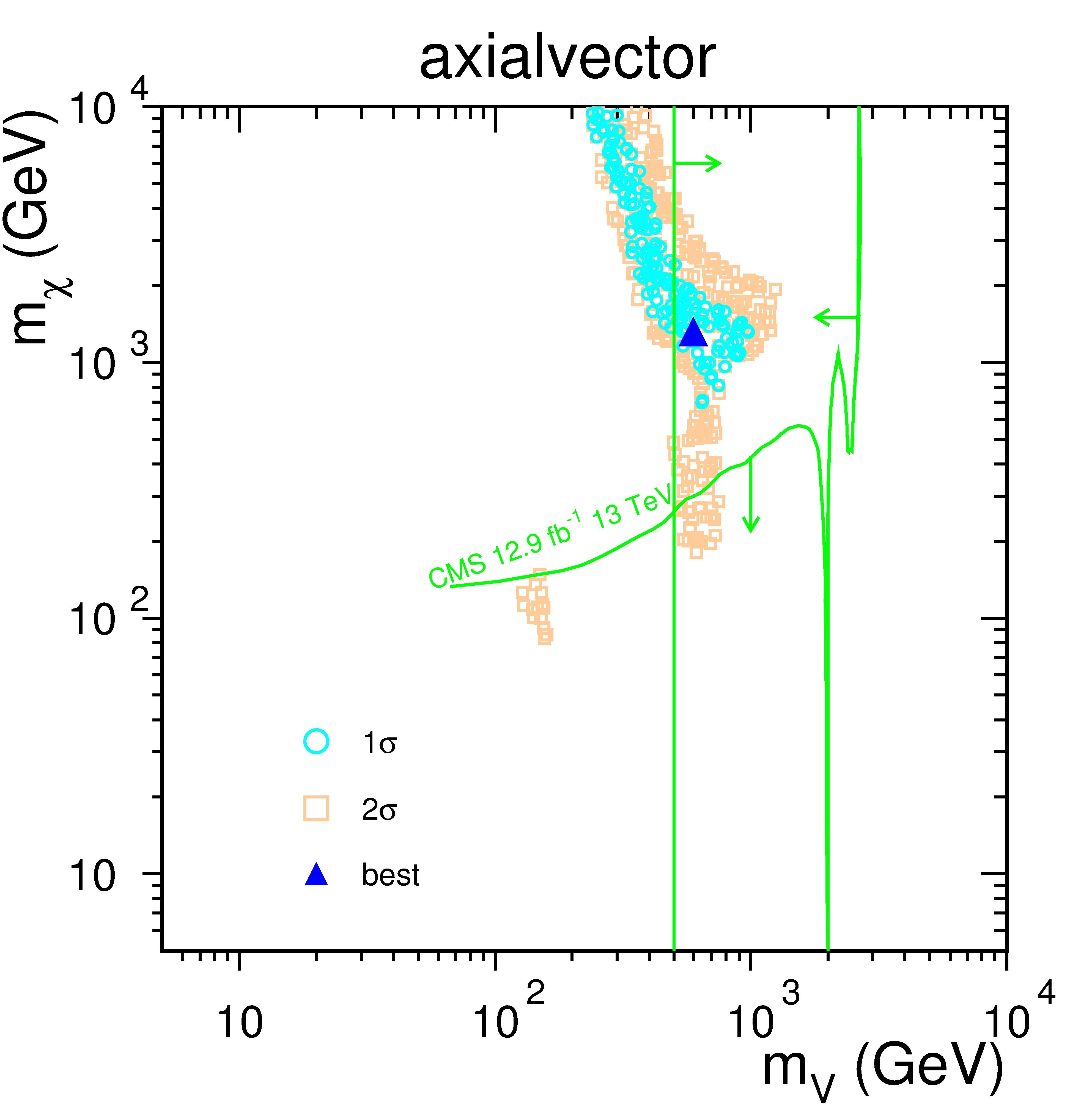}
\includegraphics[scale=1,width=7.5cm]{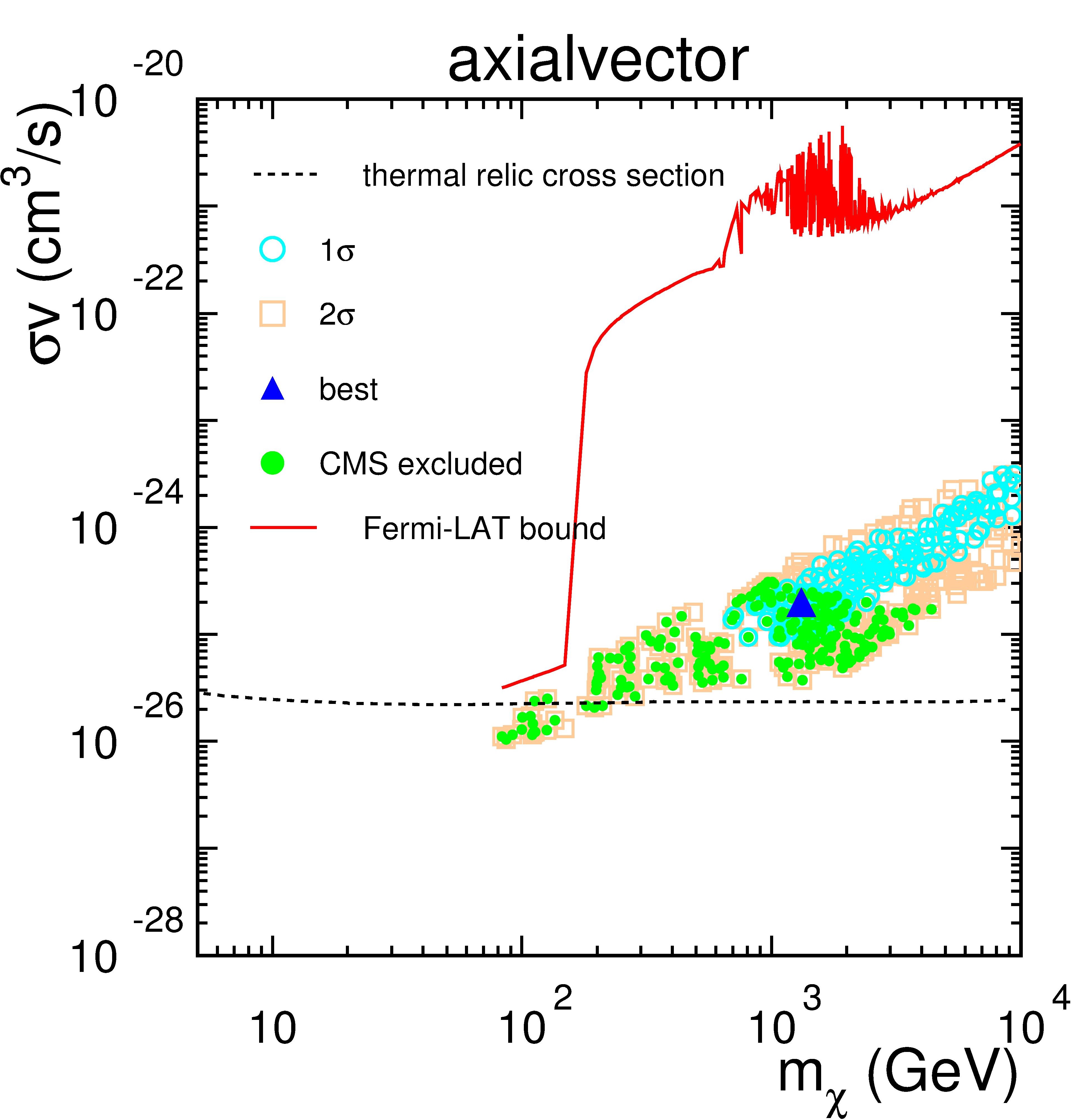}
\end{center}
\caption{Left: the AMS-02 favored region of masses ($m_\chi$ vs. $m_V$) in the simplified dark matter model with an axialvector mediator we consider. The LHC exclusion limits are from Refs.~\cite{CMS-PAS-EXO-16-037}, \cite{Khachatryan:2016ecr} and \cite{Sirunyan:2016iap}. Right: the AMS-02 favored region of cross sections ($\sigma v$ vs. $m_\chi$).}
\label{fig:region2}
\end{figure}

In the left frame of Fig.~\ref{fig:region2}, for the axialvector mediator case, we can see that the AMS-02 antiproton data favor region $m_\chi\gtrsim 700~{\rm GeV}$ and $200~{\rm GeV}\lesssim m_V\lesssim 1~{\rm TeV}$ at about $1\sigma$ confidence level. The region with $m_\chi\gtrsim 1$ TeV and $m_V\lesssim 500~{\rm GeV}$ can evade the LHC limit.

The dark matter annihilation with axialvector mediator requires the cross section in the region of $1 \times 10^{-25}$ -- $4 \times 10^{-24}$ ($1 \times 10^{-26}$ -- $4 \times 10^{-24}$) ${\rm cm}^3/{\rm s}$ at about 1(2)$\sigma$ C.L. as shown in the right plot of Fig.~\ref{fig:region2}. The LHC excludes a majority of the region below $3 \times 10^{-25} \ {\rm cm}^3/{\rm s}$, denoted by green dots. The Fermi-LAT bound does not constrain the AMS-02 favored region either.

\section{Conclusions}
\label{sec:Concl}

In this work we examine the plausibility of leptophobic dark matter annihilation contributing to the recent AMS-02 data, i.e. the antiproton flux and antiproton-to-proton ratio.
Besides the standard astrophysical cosmic ray flux prediction we include a dark matter component.  Our choice of the dark matter model is two simplified models of a Dirac fermionic dark matter, with leptophobic pseudoscalar and axialvector
mediators that couple only to SM quarks and dark matter particles.  The fluxes from standard astrophysical sources and dark matter annihilation are propagated through the Galaxy using the same set of diffusion parameters.  The propagation and injection parameters are determined by fitting the latest AMS-02 cosmic ray fluxes of nuclei.

We have shown that not only AMS-02 observations are consistent with the dark matter hypothesis within the uncertainties, but also including a dark matter contribution to the background flux gives a better fit to the data.
We also estimated the most plausible parameter regions of the dark matter parameter space in light of AMS-02 data.
The observation of antiproton prefers a dark matter (mediator) mass in the 700 GeV--5 TeV (5 GeV--10 TeV) region for the annihilation with pseudoscalar mediator and in the 700 GeV--10 TeV (200 GeV--1 TeV) region for the annihilation with axialvector mediator, respectively, at about 68\% confidence level. The AMS-02 data require an effective dark matter annihilation cross section in the region of $1 \times 10^{-25}$ -- $1 \times 10^{-24}$ ($1 \times 10^{-25}$ -- $4 \times 10^{-24}$) ${\rm cm}^3/{\rm s}$ for the simplified model with pseudoscalar (axialvector) mediator.
The LHC excludes a part of the region below thermal relic cross section for the pseudoscalar mediator model and the region with axialvector mediator mass greater than 500 GeV. The Fermi-LAT bound does not constrain the AMS-02 favored region.


\acknowledgments
We would like to thank Csaba Bal\'azs and Thomas Jacques for discussions. We also thank Qiang Yuan for helping with Galprop.
%
%
The National Computational Infrastructure (NCI), the Southern Hemisphere's fastest supercomputer, is also gratefully acknowledged.

\appendix
\section{Expressions of mediator decay widths and dark matter annihilation cross sections}

The mediator decay widths for the pseudoscalar mediator case~\cite{Boveia:2016mrp}:
\begin{eqnarray}
\Gamma_{S\to \bar{\chi}\chi} &=& {(g_{\rm DM}^{S})^2m_S\over 8\pi} \left(1-{4m_\chi^2\over m_S^2}\right)^{1/2},\\
\Gamma_{S\to \bar{q}q} &=& N_c{(g_q^{S})^2m_S\over 8\pi} {m_q^2\over v_0^2} \left(1-{4m_q^2\over m_S^2}\right)^{1/2} \ \ \ q=u,d,s,c,b,t,\\
\Gamma_{S\to gg} &=& {(g_q^{S})^2\alpha_s^2(m_S)m_S^3\over 32\pi^3 v_0^2}\left|{4m_t^2\over m_S^2}{\rm arctan}^2\left(\left({4m_t^2\over m_S^2}-1\right)^{-1/2}\right)\right|^2,\\
\Gamma_S &=& \Gamma_{S\to \bar{\chi}\chi}+ \Gamma_{S\to \bar{q}q}+\Gamma_{S\to gg}
\end{eqnarray}

The dark matter annihilation cross sections for the pseudoscalar mediator case~\cite{Arina:2014yna}:
\begin{eqnarray}
&&\sigma_{\rm ann} v(\bar{\chi}\chi\to S\to \bar{q}q) = {(g_{\rm DM}^S)^2(g_q^S)^2N_c\over (4m_\chi^2-m_S^2)^2+m_S^2\Gamma_S^2}{m_\chi^2\over 2\pi} {m_q^2\over v_0^2} \left(1-{m_q^2\over m_\chi^2}\right)^{1/2},\\
&&\sigma_{\rm ann} v(\bar{\chi}\chi\to S\to gg) = \nonumber \\
&&{(g_{\rm DM}^S)^2(g_q^S)^2\over (4m_\chi^2-m_S^2)^2+m_S^2\Gamma_S^2}{\alpha_s^2(2m_\chi)m_t^4\over 2\pi^3 v_0^2}\left|{m_t^2\over m_\chi^2}{\rm arctan}^2\left(\left({m_t^2\over m_\chi^2}-1\right)^{-1/2}\right)\right|^2,\\
&&\sigma_{\rm ann} v(\bar{\chi}\chi\to SS) = (g_{\rm DM}^S)^4{m_\chi^2(m_\chi^4-2m_\chi^2m_S^2+m_S^4)\over 24\pi (2m_\chi^2-m_S^2)^4}\left(1-{m_S^2\over m_\chi^2}\right)^{1/2}v^2,
\end{eqnarray}
where $v\simeq 10^{-3}$.

The mediator decay widths for the axialvector mediator case~\cite{Alves:2015pea}:
\begin{eqnarray}
\Gamma_{V\to \bar{\chi}\chi} &=& {(g_{\rm DM}^{A})^2m_V\over 12\pi} \left(1-{4m_\chi^2\over m_V^2}\right)^{3/2},\\
\Gamma_{V\to \bar{q}q} &=& {(g_q^{A})^2m_V\over 4\pi} \left(1-{4m_q^2\over m_V^2}\right)^{3/2} \ \ \ q=u,d,s,c,b,t,\\
\Gamma_V &=& \Gamma_{V\to \bar{\chi}\chi}+ \Gamma_{V\to \bar{q}q}
\end{eqnarray}

The dark matter annihilation cross sections for the axialvector mediator case~\cite{Alves:2015pea}:

\begin{eqnarray}
&&\sigma_{\rm ann} v(\bar{\chi}\chi\to V\to \bar{q}q) = {(g_{\rm DM}^A)^2(g_q^A)^2\over (4m_\chi^2-m_V^2)^2+m_V^2\Gamma_V^2} \left(1-{m_q^2\over m_\chi^2}\right)^{1/2}{3m_q^2(4m_\chi^2-m_V^2)^2\over 2\pi m_V^4},\nonumber \\
\\
&&\sigma_{\rm ann} v(\bar{\chi}\chi\to VV) = {(g_{\rm DM}^A)^4\over 4\pi m_\chi (2m_\chi^2-m_V^2)^2}\left(m_\chi^2-m_V^2\right)^{3/2}.
\end{eqnarray}


\end{document}